\documentclass[runningheads]{llncs}
\usepackage[T1]{fontenc}
\usepackage{graphicx}
\usepackage{todonotes}
\usepackage{pgfplots}
\pgfplotsset{compat=1.18}
\usepackage{caption}
\usepackage{subcaption}
\usepackage[margin=1in]{geometry}
\usepackage{adjustbox}

\usepackage{tabularx}
\usepackage{booktabs} % for better table lines
\usepackage{multirow}
\renewcommand{\arraystretch}{0.95}

\begin{document}
\title{The Impact of AI-Generated Solutions on Software Architecture and Productivity: Results from a Survey Study}
\titlerunning{AI: Productivity and architecture}
\author{Giorgio Amasanti\inst{1} \and
Jasmin Jahić\inst{1}\orcidID{0000-0002-8948-2960} }

\authorrunning{Amasanti and Jahić}
%\titlerunning
\institute{University of Cambridge, UK}
\maketitle              % typeset the header of the contribution
\begin{abstract}
AI-powered software tools are widely used to assist software engineers. However, there is still a need to understand the productivity benefits of such tools for software engineers. In addition to short-term benefits, there is a question of how adopting AI-generated solutions affects the quality of software over time (e.g., maintainability and extendability).

To provide some insight on these questions, we conducted a survey among software practitioners who use AI tools. Based on the data collected from our survey, we conclude that AI tools significantly increase the productivity of software engineers. However, the productivity benefits of using AI tools reduce as projects become more complex. The results also show that there are no significant negative influences of adopting AI-generated solutions on software quality, as long as those solutions are limited to smaller code snippets. However, when solving larger and more complex problems, AI tools generate solutions of a lower quality, indicating the need for architects to perform problem decomposition and solution integration. 

\end{abstract}

\section{Introduction}\label{sec:introduction}
%Recent development of AI technology and development of dedicated tools for software development has meant that AI tools are increasingly able to enhance the software development process. There already exists widespread use of AI tools by industry software developers, with 76.6\% of professional developers using or planning to use AI tools in their software development process this year according to the 2024 Stack Overflow Developer Survey \cite{stackoverflowdevelopersurvey2024}. The unknown potential for this technology raises many important questions concerning the future of software development. 

%Tools enhanced by AI are already widely used in software industry (by AI we consider Artificial Intelligence in its broadest scope, but what we mainly see in practice are tools based on Large Language Models). 
According to a study conducted by Stack Overflow in May 2024 \cite{stackoverflowdevelopersurvey2024}, 76.6\% developers are using or planning to use AI tools. Engineers use AI tools for various tasks in software engineering \cite{Chang01} \cite{Yang01} (e.g., code generation \cite{BUCAIONI2024100526},  assistance with software architecture \cite{Jahic01}). However, adopting AI in software engineering also presents challenges, as AI is a disruptive technology \cite{Jahic02}. Therefore, many practitioners (e.g., architects and managers) need to be convinced that adopting AI will bring concrete benefits without hidden "costs". %We would also like to understand influence of adopting AI solutions on software quality. %There has been a lot of research around already mentioned application areas. 
%While there has been also some research on what should be a long term frame for research when it comes to AI in software engineering \cite{manifestoCopenhagen}, we did not notice many studies that quantify holistic benefits experienced by practitioners in terms of productivity when using AI tools. 
E.g., while some AI tools certainly can provide short-term benefits (e.g., coding assistance), adopting AI-generated solutions might have negative long-term effects (e.g., adopting suboptimal AI-generated code). %This is particularity the case when AI-generates suboptimal solutions. 
In software architecture, we label suboptimal solutions as antipatterns and aim to avoid them. Hence, there is a need to understand how the adoption of AI tools (and solutions created with or generated by AI tools) influences software architecture in terms of maintainability, extendability, understandability, modifiability, cohesion, coupling, performance, and overall software partitioning. Therefore, we created a study driven by these two main research questions: i) \textbf{how AI influences the productivity of software engineers}, and ii) \textbf{ does adopting AI-generated solutions have significant negative effects on the quality of software architecture}. 

%Existing work in this area focusses mainly on instant productivity benefits, but does not pay much attention to the long-term architectural consequences of adopting AI-generated solutions. 
To gather data that can help us answer these questions, %we first created a test project during which we observed the experiences (productivity, architecture design) of an engineer in a setup with AI tools and without AI tools. This project is simple and is used as a reference point for 
we conducted a survey among 40 practitioners who use AI tools. %In this paper, we present lessons learned from the test project and the industrial survey. 
The data gathered shows that using AI tools significantly boosts productivity in the early stages of software projects and that adopting AI generated solutions does not affect software architecture in a significantly negative way. However, once there is an established code base, adopting solutions generated by AI (e.g., changing code, new design solutions) is not straightforward and can introduce significant overhead. Despite these challenges, the overwhelming majority of the surveyed professionals reported a significant increase in productivity (over 35\%) when using AI tools (for various tasks, not just coding). Therefore, \textbf{if you do not use AI tools in software engineering, you are becoming less competitive} because individuals using AI tools \textbf{\textit{significantly outperform}} those who do not use such tools for the variety of tasks in Software Development Life Cycle (SDLC).

%However, using solutions 
%changing code as project progresses by using AI tools can be challenging. On the other hand, we established that adopting AI generated code does not effect software architecture in a significantly negative way (generated code is pretty good, considering requested functionality and code quality!).

%, when they provide guidance in the exploratory phase, generate skeleton design (first software architecture), and generate first versions of code and tests for the basic features. However, changing code as project progresses by using AI tools can be challenging. On the other hand, we established that adopting AI generated code does not affect software architecture in a significantly negative way (generated code is pretty good, considering requested functionality and code quality!). However, when asking AI tools to assist with planning architecture and logical partitioning of software into different parts, one cannot take the results off the shelf as they require proper inspection. As software code base gets bigger and more complex, there is a decrease in the productivity benefits that AI tools bring. However, despite all these challenges, we learned that overwhelming majority of surveyed practitioners experience significant increase in productivity (over 35\%). Therefore, \textbf{if you are not using AI tools in software engineering, you are slowly becoming irrelevant} because our results indicate that individuals using AI tools \textbf{\textit{significantly outperform}} those who do not use them.

This paper is organised as follows. In Section \ref{sec:relatedWork}, we discuss related work. In Section \ref{sec:researchDesign}, we present our research methodology. Section \ref{sec:industrialSurvey} presents the results of the industrial survey. We conclude in Section \ref{sec:conclusion}.

\section{Related Work}\label{sec:relatedWork}

\begin{figure*}
\centering
    %\hspace*{-0.5cm}  
        \includegraphics[width=0.8\textwidth,keepaspectratio]{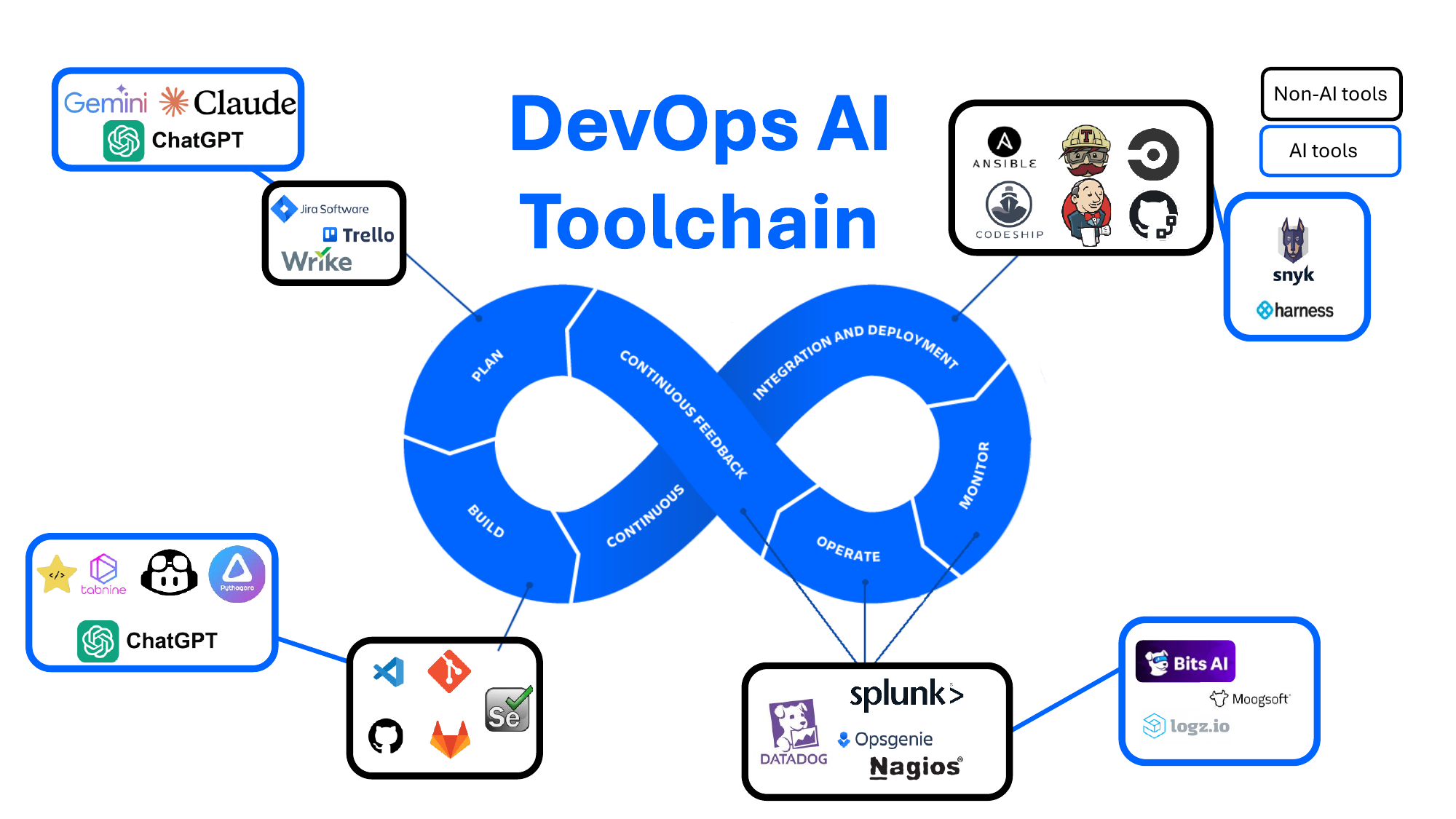}
        \caption{AI-based and AI-enabled tools in SDLC}
        \label{fig:devopsToolchainDiagram}
\end{figure*}

There exist several studies that investigate how effective AI tools are for code generation \cite{BUCAIONI2024100526} \cite{Seohyun01} \cite{OpenAI}. Engineers also use AI tools for other purposes: program repair \cite{MORADIDAKHEL2023111734}\cite{Naman01}%\cite{jiang2023impact}
\cite{zhang2022repairing}, code summarization \cite{Toufique01}\cite{Xing01}, and generating software architecture solutions \cite{Jahic01} \cite{Renjith}. In addition, there are several studies that investigate quality of solutions that these tools provide \cite{Pearce2021CanOC}, indicating that these tools suffer from low accuracy and low reliability \cite{Tufano01} \cite{Ciniselli01} \cite{Mastropaolo01} \cite{fan2023automated}. For example, how ChatGPT struggles with lengthy descriptions of problems \cite{tian2023chatgptultimateprogrammingassistant}.%Tian et al. \cite{tian2023chatgptultimateprogrammingassistant} investigated, among other things, the code generation ability of ChatGPT and found that the tool struggled on new, unseen problems and with lengthy descriptions of problems.

%, to vulnerability detection [26, 27] and providing summarization for code segments [28, 29]. 
%Experimental results in the literature demonstrate the immense potential of LLMs in the software development community. However, the overall achieved performance of these models remains relatively limited %\cite{chen2022codet}
%\cite{fan2023automated}. One area that has not been thoroughly explored, despite the increased accuracy, is the suitability of these models for more abstract tasks such as software architecture design. It is crucial to investigate the accuracy of the outputs and whether they can generate meaningful architectures that align with specifications and user requirements. 

 %The recent advances in large language models (LLM) such as GPT3 \cite{OpenAI} have led to a breakthrough in automated code generation.  

However, more recent studies \cite{nguyen2024generative} \cite{Sauvola2024}, such as the one performed by Stack Overflow \cite{stackoverflowdevelopersurvey2024} offer different perspective and indicate that using AI tools in software engineering offers productivity boosts. For example, one of the highest-performing recent AI tools, LDB (a novel AI code debugging framework using GPT-4o), can solve up to 98.2\% of the problems in HumanEval benchmark after one attempt (Pass@1)
\cite{zhong2024debuglikehumanlarge}. Bencheikh et al. \cite{bencheikh2023exploring} conducted a study comparing the quality of AI and human-written software requirements, and concluded that the AI "performs reasonably well" as a software requirements engineer. As outlined by He et al. \cite{he2024llmbasedmultiagentsystemssoftware}, the subject of much research on AI tools for software development has been focussed on the implementation of multi-agent systems. This involves utilising multiple AI entities to work together and perform separate tasks within the software engineering process, much like a human engineering team. This framework is the methodology adopted by many dedicated AI software development tools including ChatDev and MetaGPT, with a seemingly positive impact on performance in the software development process. %Hence when deciding on the tools and methodology to be used in this case study we utilised one multi-agent tool (GPT Pilot) as well as a multi-agent framework using ChatGPT.
Waseem et al. \cite{waseem2024chatgptsoftwaredevelopmentbot} conducted a case study similar to ours investigating the viability of ChatGPT as a software engineering tool. Across the software development process, the students involved in the study reported the AI was effective at: identifying requirements, streamlining development via code generation, bug detection and architectural decision making. However, it fell short on some metrics including: "design oversimplification", "creativity hindrance", "constraint conflict", and "architectural expertise neglect". Compared to the existing work, we focus on productivity and quality of software design, when adopting AI-generated solutions.

\section{Research Study Design}
\label{sec:researchDesign}

\renewcommand{\arraystretch}{1.1}
\begin{table}[htbp]
\centering
\caption{GQM Mapping: Tetris Project Study}
\begin{tabular}{|p{2.2cm}|p{5.8cm}|p{7.8cm}|}
\hline
\textbf{Goal} & \textbf{Question} & \textbf{Metric} \\
\hline

\multirow{3}{=}{\textbf{G1: Identify AI tools for SDLC}} 
& \textbf{Q1.1:} Which AI tools are currently available across SDLC stages? 
& \textbf{M1.1:} Tool name and SDLC stage(s) supported; context of use. \\
\cline{2-3}
& \textbf{Q1.2:} Which widely used SDLC tools have AI-enabled features? 
& \textbf{M1.2:} Tool name and vendor; list of AI subcomponents/features. \\
\cline{2-3}
& \textbf{Q1.3:} Are there integrated AI-powered SDLC toolchains? 
& \textbf{M1.3:} Toolchain structure; SDLC stages covered. \\
\hline

\multirow{2}{=}{\textbf{G2: Analyze suitability of AI for SDLC stages}} 
& \textbf{Q2.1:} What tasks are commonly assisted or automated by AI? 
& \textbf{M2.1:} Task–tool mapping; automation level. \\
\cline{2-3}
& \textbf{Q2.2:} Which SDLC stages are difficult to delegate to AI? 
& \textbf{M2.2:} Stages flagged as unsuitable; challenges reported per stage. \\
\hline

\multirow{3}{=}{\textbf{G3: Assess productivity impact of AI tools}} 
& \textbf{Q3.1:} Does AI reduce time required for specific tasks? 
& \textbf{M3.1:} Estimated time saved per task/project. \\
\cline{2-3}
& \textbf{Q3.2:} How often do AI outputs require rework? 
& \textbf{M3.2:} Percentage of AI outputs needing edits; rework frequency. \\
\cline{2-3}
& \textbf{Q3.3:} How do developers assess productivity with AI? 
& \textbf{M3.3:} Self-rated productivity scores with/without AI. \\
\hline

\multirow{3}{=}{\textbf{G4: Examine architectural implications of adopting AI-generated solutions}} 
& \textbf{Q4.1:} How correct and adequate are AI-generated solutions? 
& \textbf{M4.1:} Percentage of AI outputs used without modification; number of bugs. \\
\cline{2-3}
& \textbf{Q4.2:} Are AI-generated solutions harder to modify or test? 
& \textbf{M4.2:} Time to understand/edit/integrate AI code; difficulty rating. \\
\cline{2-3}
& \textbf{Q4.3:} How does AI-generated code affect quality of architecture? 
& \textbf{M4.3:} System consistency; integration issues; maintainability issues; extendibility issues; understandability issues; cohesion and coupling issues; logical partitioning issues. \\
\hline

\end{tabular}
\label{table:gqmTetrisProjectStudy}
\end{table}

Our study is composed out of two substudies. In the first sub-study, we created an instance of the classic Tetris videogame with the help of AI tools. We used this sub-study to scan the market for available AI tools and to get initial results of how these tools behave when integrated into Software Development Life Cycle (SDLC). We used the conclusions and experiences from the first substudy to setup a survey with industry professionals (second substudy). To structure our study, we used Goal-Question-Metric (GQM) approach~\cite{rombach}~\cite{Basili}. For the design of the survey, we followed the established guidelines~\cite{kasunic2005designing}. To analyse the data gathered in the survey, we applied the principles of grounded theory using open and axial coding.% to identify the key emerging themes and the relationships among them.

\subsection{Tetris Project Study: Design}\label{sec:researchDesign:tetrisDesign}

\begin{table}[htbp]
\renewcommand{\arraystretch}{1.15}
\centering
\caption{Mapping GQM Structure to Tetris Project Results}
\begin{tabular}{|p{0.6cm}|p{0.8cm}|p{0.8cm}|p{13.5cm}|}
\hline
\textbf{G} & \textbf{Q} & \textbf{M} & \textbf{Finding / Observation} \\
\hline

G1 & Q1.1 & M1.1 & Identified: ChatGPT, StarCoder, Llama (code generation); GitHub Copilot, Cursor IDE (IDE support); Snyk, Harness (CI/CD); Datadog, Splunk, Logz.io, Moogsoft (ops). \\
\cline{2-4}
   & Q1.2 & M1.2 & Features observed in tools like GPT Pilot, ChatDev, Devin, MetaGPT, AutoDev, with multi-stage AI support. \\
\cline{2-4}
   & Q1.3 & M1.3 & Tools like GPT Pilot and ChatDev integrate multiple SDLC stages including requirements, coding, and review. \\
\hline

G2 & Q2.1 & M2.1 & AI is effective in early implementation and boilerplate generation. Less effective in bug fixing and integration. \\
\cline{2-4}
   & Q2.2 & M2.2 & Bug fixing and code modification stages require multiple prompts; productivity drops as complexity increases. \\
\hline

G3 & Q3.1 & M3.1 & AI improves productivity by up to 35\%, especially during early-stage development. However, long response times from AI tools have a significant negative impact on productivity. As code complexity increases, AI response times also increase (e.g., up to 10 minutes per prompt), reducing productivity gains.\\
\cline{2-4}
   & Q3.2 & M3.2 & Typically 2–3 prompt iterations are needed; the need for rework increases with code complexity, as AI-generated solutions in complex software tend to introduce bugs. \\
\cline{2-4}
   & Q3.3 & M3.3 & Asking for AI assistance on small, focused problems maximizes productivity. Attempting to solve large problems in a single prompt often leads to errors. Manual editing is often faster for integration and bug fixes (AI often requires multiple prompts). Humans comprehend the code better than AI. \textbf{}\\
\hline

G4 & Q4.1 & M4.1 & AI-generated code is almost always syntactically correct. AI rarely produces functionally adequate results on first attempt and the functional adequacy of generated code decreases as complexity grows. \\
\cline{2-4}
   & Q4.2 & M4.2 & The logical style of AI-generated code—such as naming consistency, modifiability, and line length—is generally comparable to human-written code, contributing to good readability and understandability. AI requires more time than humans for editing and debugging as it often requires multiple prompts. Integrating AI-generated code into an existing codebase can require substantial manual effort, such as renaming variables or adapting function signatures and module interfaces.\\
\cline{2-4}
   & Q4.3 & M4.3 & Large AI-generated code has poor modularity, maintainability, extensibility, and understandability due to illogical structure. Small snippets yield high cohesion, low coupling and a clear separation of concerns, resembling human-written code in architectural quality. Smaller AI-generated code snippets also match human-written code in terms of maintainability, including
structural quality and clarity. AI-generated code demonstrates performance characteristics comparable to those of human-written code, with no significant overhead observed.\\
\hline

\end{tabular}
\label{table:gqm-results-tetris}
\end{table}

Table~\ref{table:gqmTetrisProjectStudy} presents the detailed design (using the GQM approach) for our substudy in which we created an instance of a classical Tetris game. In this substudy, we aimed to investigate how AI tools behave when integrated in SDLC compared to the work done without using AI tools. To conduct the study, we sought a junior \textit{software engineer} with proficiency in at least one programming language, and an experienced software engineering professional to serve as the \textit{project commissioner}, responsible for defining the business architecture and high-level requirements. The development %of the Tetris instance 
was organised using the Scrum framework~\footnote{\url{https://www.scrum.org/resources/what-scrum-module}}.

\subsection{Tetris Project Study: Results}\label{sec:researchDesign:tetrisResults}

%The \textit{project commissioner} collected a list of the most prominent commercial AI-based tools through a simple Google search. Based on the experiments with those tools, \textit{project commissioner} created a toolchain for the project study. 
The substudy was conducted over a period of six weeks. In the first half of the study, the \textit{software engineer} scanned the market for available AI-tools (Google search, systematic literature reviews, case studies, and reports on using AI in software engineering) and used those tools to experiment with setting up tool chains to cover stages of the Software Development Life Cycle (SDLC). In the second half of the study, over a period of the next three weeks (comprising three one-week sprints), the \textit{software engineer}, guided by instructions from the \textit{project commissioner}, developed three parallel versions of the Tetris game, all three based on the same set of requirements. Each version followed a different technical setup. In \textbf{version a)}, the \textit{software engineer} did not use any AI tools. In \textbf{version b)}, the engineer used GPT Pilot—a terminal-based code planning and generation tool leveraging GPT-4o-mini-2024-07-18 via OpenAI’s API—and ChatGPT running on GPT-4o-2024-05-13. In \textbf{version c)}, the \textit{software engineer} employed multiple ChatGPT instances (GPT-4o-2024-05-13) to simulate distinct roles: (i) project manager—configures other roles; (ii) requirements engineer—generates refined requirements; (iii) software architect—designs the base software architecture; (iv) developer agent—writes and debugs code; and (v) quality assurance (QA)/quality control (QC) engineer—evaluates and provides feedback on code quality. The \textit{software engineer} manually copied inputs and outputs between these roles to advance development. The results of the study are summirized in Table \ref{table:gqm-results-tetris}, and the full and raw results, as well as the code, are available online: github.com/AItoolsSE/CAIN-AI-documents; github.com/AItoolsSE/CAIN-AI-approach-A; github.com/AItoolsSE/CAIN-AI-approach-B; github.com/AItoolsSE/CAIN-AI-approach-C %\footnote{https://github.com/AItoolsSE/CAIN-AI-documents} \footnote{https://github.com/AItoolsSE/CAIN-AI-approach-A} \footnote{https://github.com/AItoolsSE/CAIN-AI-approach-B} \footnote{https://github.com/AItoolsSE/CAIN-AI-approach-C}. 

\subsection{Survey with Industry Professionals: Design}\label{sec:researchDesign:surveyDesign}

We used conclusions from the Tetris Project Study (Section \ref{sec:researchDesign:tetrisResults}) to create a structure of the survey with industry professionals. Table~\ref{table:gqm-ai-survey} presents the detailed study design using the GQM approach. To conduct the study, we sought industry professionals who are already using some AI tools in their daily work. We aimed for 30 participants, with a variety considering the institution size, years of experience, roles in SDLC, and geography locations. We invited professionals from our network and shared the survey on LinkedIn.

\renewcommand{\arraystretch}{1.1}
\begin{table}[htbp]
\centering
\caption{GQM: AI Tools in Software Engineering – Productivity and Architecture}
\begin{tabular}{|p{2.2cm}|p{6.0cm}|p{7.8cm}|}
\hline
\textbf{Goal} & \textbf{Question} & \textbf{Metric} \\
\hline

\multirow{2}{=}{\textbf{G1: Identify AI tools used across the SDLC}} 
& \textbf{Q1.1:} Which tasks in the SDLC are supported by AI tools? 
& \textbf{M1.1:} List of SDLC tasks (e.g., code generation, test generation, CI/CD, bug fixing) \\
\cline{2-3}
& \textbf{Q1.2:} Which AI tools are used in daily development work? 
& \textbf{M1.2:} Tool names selected (e.g., Copilot, ChatGPT) \\
\hline

\multirow{4}{=}{\textbf{G2: Assess the productivity benefits\\ of using AI tools}} 
& \textbf{Q2.1:} How do software engineers perceive productivity improvements with AI tools? 
& \textbf{M2.1:} Self-reported productivity impact relative to a reference of up to 35\% improvement (ordinal scale: much lower to much higher) \\
\cline{2-3}
& \textbf{Q2.2:} At which SDLC stage is AI usage most beneficial? 
& \textbf{M2.2:} Agreement on usefulness in early implementation (e.g., generating skeleton code) \\
\cline{2-3}
& \textbf{Q2.3:} What is the most effective way to apply AI tools during development? 
& \textbf{M2.3:} Agreement on effectiveness of decomposition (e.g., prompts on smaller problems yield better results) \\
\cline{2-3}
& \textbf{Q2.4:} What factors hinder productivity when using AI tools? 
& \textbf{M2.4:} Negative impact of processing time; increased latency with code size; bug frequency in larger outputs; average number of prompt repetitions; time to fix issues; manual integration effort (e.g., renaming variables, adapting interfaces) \\
\hline

\multirow{8}{=}{\textbf{G3: Examine architectural challenges of adopting AI-generated code}} 
& \textbf{Q3.1:} How syntactically correct is AI-generated code? 
& \textbf{M3.1:} Frequency of syntactic correctness (Always, Often, Sometimes, Rarely) \\
\cline{2-3}
& \textbf{Q3.2:} How functionally adequate is AI-generated code on first attempt? 
& \textbf{M3.2:} Self-reported success rate; prompt repetition; influence of complexity \\
\cline{2-3}
& \textbf{Q3.3:} Does large-scale AI-generated code reduce logical organisation and architectural quality? 
& \textbf{M3.3:} Agreement with poor structure in large AI-generated blocks (low maintainability, extendability) \\
\cline{2-3}
& \textbf{Q3.4:} Does the use of AI increase architectural erosion (e.g., cohesion, coupling)? 
& \textbf{M3.4:} Agreement on architectural erosion; depends on size and structure of AI-generated code \\
\cline{2-3}
& \textbf{Q3.5:} Is AI-generated code comparable in performance to human-written code? 
& \textbf{M3.5:} Agreement with performance quality (e.g., minimal overhead) \\
\cline{2-3}
& \textbf{Q3.6:} Does AI-generated code affect maintainability and modifiability? 
& \textbf{M3.6:} Agreement with difficulty updating/extending AI-generated code \\
\cline{2-3}
& \textbf{Q3.7:} What kinds of tests are generated by AI tools? 
& \textbf{M3.7:} Presence of advanced testing patterns (e.g., mocks, pixel-level testing) \\
\cline{2-3}
& \textbf{Q3.8:} How reliable are AI-generated tests? 
& \textbf{M3.8:} Initial test failure rate  \\
\hline

\end{tabular}
\label{table:gqm-ai-survey}
\end{table}

\section{The Impact of AI-Generated Solutions on Software Architecture
and Productivity}\label{sec:industrialSurvey}

\subsection{Demographics}\label{sec:industrialSurvey:demographics}

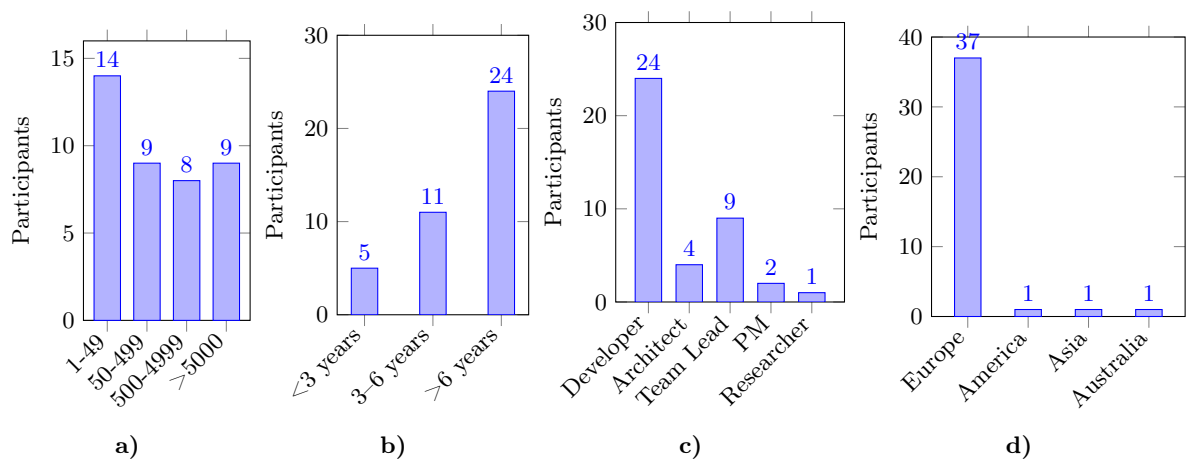
\begin{figure}[htbp]
    \centering
    \begin{adjustbox}{max width=\textwidth}
    \begin{tabular}{cccc}
        % Chart 1: Institution size
        \begin{tikzpicture}
        \begin{axis}[
            ybar,
            ylabel={Participants},
            symbolic x coords={1-49, 50-499, 500-4999, >5000},
            xtick=data,
            xticklabel style={rotate=45, anchor=east},
            nodes near coords,
            ymin=0,
            ymax=16,
            bar width=10pt,
            enlarge x limits=0.2,
            width=0.23\textwidth,
            height=5.3cm,
        ]
        \addplot coordinates {
            (1-49,14)
            (50-499,9)
            (500-4999,8)
            (>5000,9)
        };
        \end{axis}
        \end{tikzpicture}
        &
        % Chart 2: Experience
        \begin{tikzpicture}
        \begin{axis}[
            ybar,
            ylabel={Participants},
            symbolic x coords={<3 years, 3–6 years, >6 years},
            xtick=data,
            xticklabel style={rotate=45, anchor=east},
            nodes near coords,
            ymin=0,
            ymax=30,
            bar width=10pt,
            enlarge x limits=0.2,
            width=0.25\textwidth,
            height=5.3cm,
        ]
        \addplot coordinates {
            (<3 years,5)
            (3–6 years,11)
            (>6 years,24)
        };
        \end{axis}
        \end{tikzpicture}
        &
        % Chart 3: Current role
        \begin{tikzpicture}
        \begin{axis}[
            ybar,
            ylabel={Participants},
            symbolic x coords={Developer, Architect, Team Lead, PM, Researcher},
            xtick=data,
            xticklabel style={rotate=45, anchor=east},
            nodes near coords,
            ymin=0,
            ymax=30,
            bar width=10pt,
            enlarge x limits=0.2,
            width=0.28\textwidth,
            height=5.3cm,
        ]
        \addplot coordinates {
            (Developer,24)
            (Architect,4)
            (Team Lead,9)
            (PM,2)
            (Researcher,1)
        };
        \end{axis}
        \end{tikzpicture}
        &
        % Chart 4: Continent
        \begin{tikzpicture}
        \begin{axis}[
            ybar,
            ylabel={Participants},
            symbolic x coords={Europe, America, Asia, Australia},
            xtick=data,
            xticklabel style={rotate=45, anchor=east},
            nodes near coords,
            ymin=0,
            ymax=40,
            bar width=10pt,
            enlarge x limits=0.2,
            width=0.3\textwidth,
            height=5.3cm,
        ]
        \addplot coordinates {
            (Europe,37)
            (America,1)
            (Asia,1)
            (Australia,1)
        };
        \end{axis}
        \end{tikzpicture}
        \\
        \textbf{a)} & \textbf{b)} & \textbf{c)} & \textbf{d)}
    \end{tabular}
    \end{adjustbox}
    \caption{Participants by a) institution size, b) experience (PM - Project Manager), c) role, and d) office location}
    \label{fig:participant-demographics}
\end{figure}

We conducted the survey in November 2024 %among our personal contacts in the industry 
and received responses from \textbf{\textit{40 participants}} (\textbf{\textit{the raw survey results are available online}}\footnote{https://github.com/AItoolsSE/CAIN-AI-documents}) %from 4 continents (Europe, America, Asia and Australia), 
from 13 different countries and companies of various sizes (Figure \ref{fig:participant-demographics}-a). Most of the survey participants were senior professionals (Figure \ref{fig:participant-demographics}-b) coming from Europe. % (24, 60\%). %We also had 9 team leads, 4 software architects, 1 researcher, and 1 project manager. We have a very good representation of companies of various sizes (Figure \ref{fig:experienceRole}-a) and the survey participants are quite experienced as illustrated in Figure \ref{fig:experienceRole}-b, with 60\% of them having over 6 years of experience in software engineering.

\subsection{Results}\label{sec:industrialSurvey:results}

\begin{table}[htbp]
\centering
\renewcommand{\arraystretch}{1.1}
\caption{Open Coding Results from Survey Responses}
\begin{tabular}{|p{3.0cm}|p{12.5cm}|}
\hline
\textbf{Code} & \textbf{Example Excerpts} \\
\hline
Code generation assistance & "Generate small snippets of code", "Generate code for full applications", "Generate CI/CD scripts", "Generate tests" \\
\hline
Test generation & "Generate tests", "Did not use AI for tests", "Tests are not based on requirements"\\
\hline

Bug fixing & "Ask AI tools to fix bugs" \\
\hline
Testing reliability concerns & "AI tends to generate tests to make code succeed", "Waste of time for significant features", “Tests not based on requirements” \\
\hline
Code modification & "Ask AI tools to modify existing code", "Refactoring code" \\
\hline
Productivity improvement perception (35\% as a reference value) & "Higher productivity improvements", "Lower productivity", “The same productivity improvements”, “AI helps with boilerplate”, "Blindly trusting AI can hurt productivity" \\
\hline
Prompt effectiveness & "Prompts really matter", "AI forgets instructions after a few prompts" \\
\hline
Integration effort & "Manual integration like renaming variables", "Needs human revision", "Needs interface adaptation", “Works best when the problem is well explained”, “Helpful for generating skeleton code”, “Saves time for documentation lookup” \\
\hline
Context misunderstanding & "AI doesn't understand my end goal", "Lacks context", "Only works well if problem is well explained" \\
\hline
Useful for early-stage work & "Brainstorming before writing code", "Useful for generating skeleton code", "Saves time on documentation lookup" \\
\hline
Limits in complex tasks & “Not good for system architecture design”, “Needs careful revision”, “Performance optimization requires human touch” \\
\hline
Architectural erosion concerns & "Reduces maintainability", "Low cohesion", "Poor structure in large AI-generated blocks" \\
\hline
\end{tabular}
\label{table:open-coding}
\end{table}

As the first step in grounded theory, we performed open-coding analysis (Table \ref{table:open-coding}) on the raw qualitative data, breaking them into discrete parts and labelling them with codes that represent the underlying concepts, actions, and concerns. The purpose was to identify patterns, recurring themes, or meaningful units in the data without imposing pre-existing categories. To build meaningful connections between data fragments in the survey, we applied axial coding (Table \ref{table:axial-coding-expanded}) on the results of the open-coding analysis. Finally, we mapped these insights to our GQM table (Table \ref{table:survey-gqm-mapping}).

\begin{table}[htbp]
\centering
\renewcommand{\arraystretch}{1.2}
\caption{Axial Coding Categories, Subcategories, and Relationships}
\begin{tabular}{|p{2.5cm}|p{9.5cm}|p{3.5cm}|}
\hline
\textbf{Category} & \textbf{Subcategories} & \textbf{Relationship} \\
\hline
AI Support Areas in SDLC &
- Code support: generation (snippets, full apps), modification, refactoring \newline
- Testing support: generation of tests (often shallow or unreliable) \newline
- Debugging: bug fixing and error correction &
AI is primarily integrated in early implementation, debugging, and testing phases \\
\hline
Perceived Productivity Impacts &
- Positive: higher productivity in early development and boilerplate tasks \newline
- Neutral/Negative: same or lower productivity in complex or misunderstood contexts \newline
- Influencing factors: prompt effectiveness, user experience, task complexity &
Gains arise in repetitive/exploratory work; drop when context is misunderstood \\
\hline
AI Limitations &
- Context comprehension: poor handling of high-level architectural goals \newline
- Integration effort: requires manual variable renaming, interface adaptation \newline
- Reliability issues: especially for testing and performance tuning &
Limitations affect productivity and architecture quality, especially for complex systems \\
\hline
Prompting Skills and Experience &
- Skilled users: understand tool boundaries and maximize utility \newline
- Inexperienced users: risk overreliance or misuse \newline
- Prompt engineering: critical for effective AI use &
Prompting capability strongly mediates the benefit obtained from AI tools \\
\hline
Tool and Use Context &
- Tool choice: Claude, ChatGPT, Copilot, Gemini, etc. \newline
- Context: variations across roles, company size, maturity of AI integration &
Effectiveness and perception depend on tooling ecosystem and role-based context \\
\hline
\end{tabular}
\label{table:axial-coding-expanded}
\end{table}

\begin{figure}[ht]
\centering
\begin{tikzpicture}
\begin{axis}[
    ybar,
    bar width=0.5cm,
    width=17cm,
    height=4cm,
    ylabel={Participants},
    symbolic x coords={
        {Generate small\\snippets of\\ code},
        {Ask AI tools \\to modify \\existing code},
        {Ask AI tools\\ to fix bugs},
        {Generate\\tests},
        {Generate code for\\  large parts\\ of applications},
        {Generate\\CI/CD\\ scripts}
    },
    xtick=data,
    x tick label style={align=center, rotate=0},
    nodes near coords,
    nodes near coords align={vertical},
    ymin=0,
    enlarge y limits=false,
    axis on top=true
]
\addplot coordinates {
    ({Generate small\\snippets of\\ code},38)
    ({Ask AI tools \\to modify \\existing code},26)
    ({Ask AI tools\\ to fix bugs},19)
    ({Generate\\tests},17)
    ({Generate code for\\  large parts\\ of applications},8)
    ({Generate\\CI/CD\\ scripts},6)
};
\end{axis}
\end{tikzpicture}
\caption{Usage of AI for Coding Tasks}
\label{figure:SDLCstages}
\end{figure}
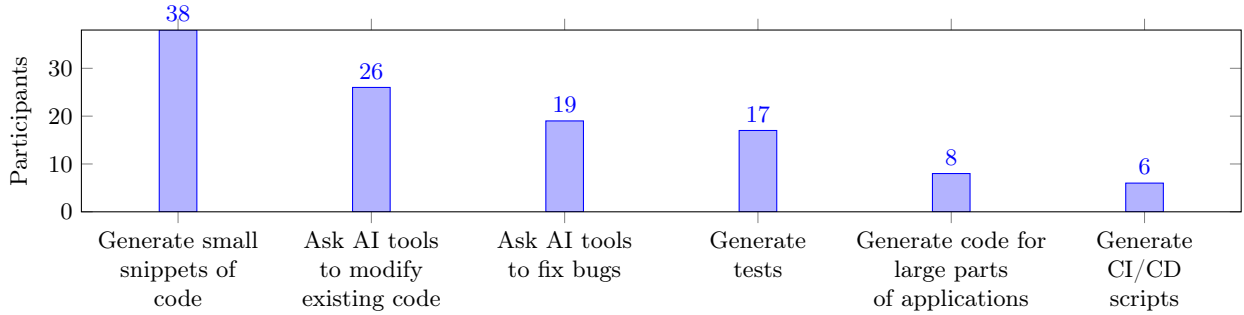

\begin{figure}[ht]
\centering
\begin{tikzpicture}
\begin{axis}[
    ybar,
    bar width=0.4cm,
    width=14cm,
    height=4cm,
    ylabel={Participants},
    symbolic x coords={
        ChatGPT, Copilot, Claude AI, grammarly GO, GPT Pilot,
        Cursor IDE, Gemini, Amazon Q, CodeWhisperer,
        Open-source projects, In-house tools
    },
    xtick=data,
    x tick label style={rotate=45, anchor=east},
    nodes near coords,
    nodes near coords align={vertical},
    ymin=0,
    enlarge y limits=false,
    axis on top=true
]
\addplot coordinates {
    (ChatGPT,32)
    (Copilot,21)
    (Claude AI,5)
    (grammarly GO,3)
    (GPT Pilot,3)
    (Cursor IDE,3)
    (Gemini,2)
    (Amazon Q,1)
    (CodeWhisperer,1)
    (Open-source projects,1)
    (In-house tools,1)
};
\end{axis}
\end{tikzpicture}
\caption{AI tools used by the survey participants}
\label{figure:aiTools}
\end{figure}
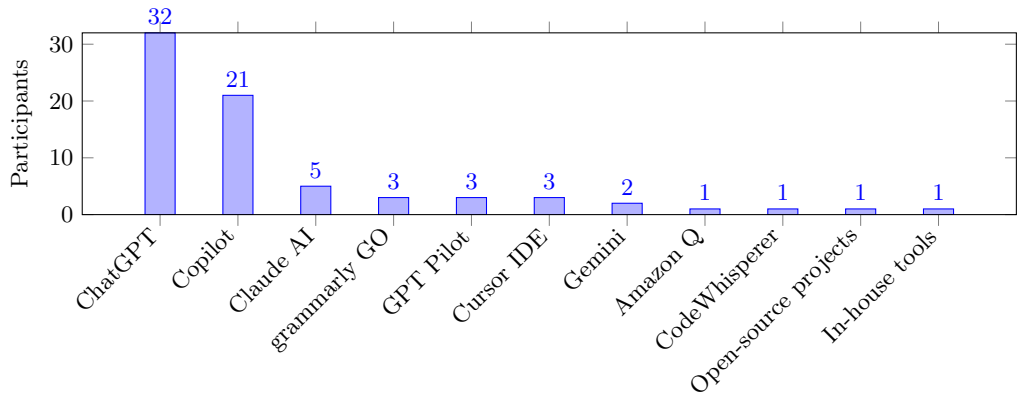

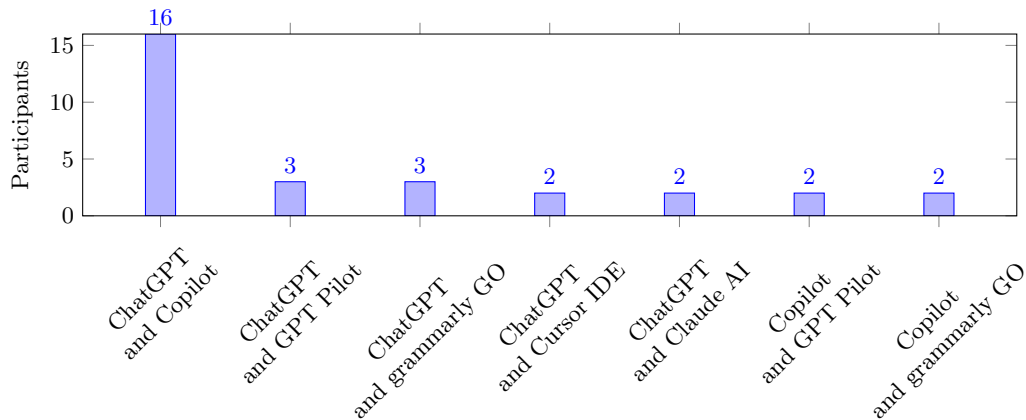
\begin{figure}[ht]
\centering
\begin{tikzpicture}
\begin{axis}[
    ybar,
    bar width=0.4cm,
    width=14cm,
    height=4cm,
    ylabel={Participants},
    symbolic x coords={
        {ChatGPT\\and Copilot},
        {ChatGPT\\and GPT Pilot},
        {ChatGPT\\and grammarly GO},
        {ChatGPT\\and Cursor IDE},
        {ChatGPT\\and Claude AI},
        {Copilot\\and GPT Pilot},
        {Copilot\\and grammarly GO}
    },
    xtick=data,
    x tick label style={rotate=45, align=center},
    nodes near coords,
    nodes near coords align={vertical},
    ymin=0,
    enlarge y limits=false,
    axis on top=true
]
\addplot coordinates {
    ({ChatGPT\\and Copilot},16)
    ({ChatGPT\\and GPT Pilot},3)
    ({ChatGPT\\and grammarly GO},3)
    ({ChatGPT\\and Cursor IDE},2)
    ({ChatGPT\\and Claude AI},2)
    ({Copilot\\and GPT Pilot},2)
    ({Copilot\\and grammarly GO},2)
};
\end{axis}
\end{tikzpicture}
\caption{Tool combinations used by the survey participants}
\label{figure:combinationsOfAITools}
\end{figure}

From the results obtained by the application of grounded theory, we conclude the following. The most common uses of AI considering SDLC are to generate small code snippets, modify existing code, fix bugs, and generate tests cases (Figure \ref{figure:SDLCstages}). The most popular AI tools in the industry are ChatGPT and Copilot (Figure \ref{figure:aiTools}). We observed that 50\% of the survey participants who use ChatGPT also use Copilot (Figure \ref{figure:combinationsOfAITools}).

Measuring \textit{\textbf{productivity}} in software engineering is not easy \cite{Suzana01}. Therefore, we decided to ask for overall \textit{\textbf{subjective feeling of productivity}} that each survey participant has when using AI tools. In that subjective context, we did try to understand if AI is more helpful with some tasks and if it introduces overhead in others. Based on our experience from the test project and  based on the experiences reported in related work, we decided to differentiate between i) applying AI tools to create a fresh code base and ii) applying AI to an already established project. The results of the survey are presented in a way that reflects this differentiation. 

When considering 35\% increase in productivity as a reference value (as observed in our test project), the results (Figure \ref{figure:productivityImprovements}) show that 79\% of the surveyed participants experience the same, higher or much higher productivity when using AI tools. It is important to note that 21\% of the participants experienced lower improvements in productivity or no improvements in productivity at all. We analysed the results of those that indicated lower productivity, but could not find any correlation with other answers. These participants use ChatGPT and Copilot (the same as those who experienced increase in productivity), and use these tools for similar tasks. The only interesting thing here might be that all the juniors (below 3 years of work experience) said that AI tools increase their productivity (none of the juniors was among 21\% of the participants who experienced lower improvements in productivity or no improvements in productivity). Majority of the survey participants agree that the highest productivity increase with AI tools can be achieved to generate a skeleton of an application and generate the basic features of the application (Figure \ref{figure:momentToApplyAITools}).

\begin{table}[htbp]
\centering
\renewcommand{\arraystretch}{1.2}
\caption{Survey results: Mapping of analysis results to GQM elements}
\begin{tabular}{|p{0.6cm}|p{0.8cm}|p{0.8cm}|p{13.5cm}|}
\hline
\textbf{G} & \textbf{Q} & \textbf{M} & \textbf{Finding/Observation} \\
\hline
\multirow{2}{*}{G1} & Q1.1 & M1.1 & AI tools are used across SDLC for code generation, bug fixing, test generation, CI/CD, and documentation support. \\
\cline{2-4}
& Q1.2 & M1.2 & Tools reported include ChatGPT, Claude, Copilot, Cursor IDE, and others with varying preferences based on user experience. \\
\hline
\multirow{4}{*}{G2} & Q2.1 & M2.1 & Perceived productivity gains range from much higher to none, depending on the SDLC task, developer experience, and task complexity. There is a strong indication of a significant increase in engineering performance productivity (over 35\% increase in productivity).\\
\cline{2-4}
& Q2.2 & M2.2 & AI is considered most beneficial in early implementation stages (e.g., boilerplate, skeleton code). \\
\cline{2-4}
& Q2.3 & M2.3 & Smaller, well-scoped prompts are more effective; prompting skill is a key enabler of productivity. \\
\cline{2-4}
& Q2.4 & M2.4 & Barriers include manual integration effort, low test reliability, prompt iteration overhead, and poor handling of complex tasks. For smaller code snippets and smaller problems, the challenges are minimal.\\
\hline
\multirow{8}{*}{G3} & Q3.1 & M3.1 & AI-generated code is often syntactically correct but varies in quality depending on prompt clarity and task complexity. \\
\cline{2-4}
& Q3.2 & M3.2 & First-attempt functional adequacy is inconsistent; requires iterative prompting, especially for complex problems. \\
\cline{2-4}
& Q3.3 & M3.3 & Larger AI-generated blocks reduce architectural clarity and maintainability; structure quality degrades. \\
\cline{2-4}
& Q3.4 & M3.4 & Participants noted signs of architectural erosion (primarily lower cohesion) in AI-generated code for larger problems and larger code bases. These are minimal for smaller problems and when integrating smaller generated code snippets.\\
\cline{2-4}
& Q3.5 & M3.5 & Performance of AI-generated code is acceptable for simpler cases. It can be affected as the problem size increases. \\
\cline{2-4}
& Q3.6 & M3.6 & Maintainability can suffer due to lack of structure, contextual gaps, and ambiguous logic. \\
\cline{2-4}
& Q3.7 & M3.7 & Tests generated are often superficial; lack coverage based on requirements or edge cases. \\
\cline{2-4}
& Q3.8 & M3.8 & High initial failure rate of tests noted; AI tends to generate tests to pass existing code, not validate correctness. \\
\hline
\end{tabular}
\label{table:survey-gqm-mapping}
\end{table}

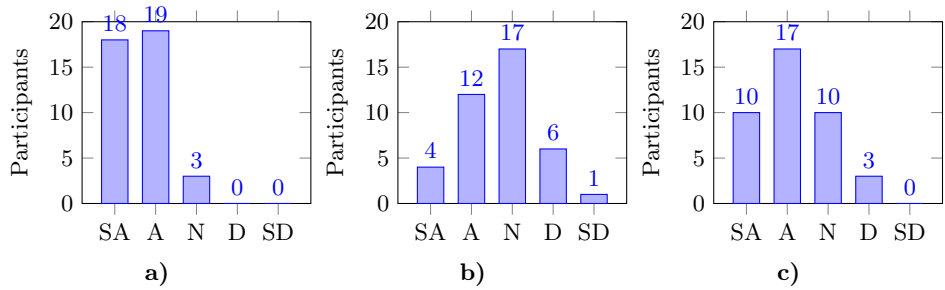
\begin{figure}[htbp]
    \centering
    \begin{adjustbox}{max width=\textwidth}
    \begin{tabular}{ccc}
        % Chart a
        \begin{tikzpicture}
        \begin{axis}[
            ybar,
            ylabel={Participants},
            symbolic x coords={SA, A, N, D, SD},
            xtick=data,
            xticklabel style={rotate=0},
            nodes near coords,
            ymin=0,
            ymax=20,
            bar width=10pt,
            enlarge x limits=0.2,
            width=0.28\textwidth,
            height=4cm,
        ]
        \addplot coordinates {
            (SA,18)
            (A,19)
            (N,3)
            (D,0)
            (SD,0)
        };
        \end{axis}
        \end{tikzpicture}
        &
        % Chart b
        \begin{tikzpicture}
        \begin{axis}[
            ybar,
            ylabel={Participants},
            symbolic x coords={SA, A, N, D, SD},
            xtick=data,
            xticklabel style={rotate=0},
            nodes near coords,
            ymin=0,
            ymax=20,
            bar width=10pt,
            enlarge x limits=0.2,
            width=0.28\textwidth,
            height=4cm,
        ]
        \addplot coordinates {
            (SA,4)
            (A,12)
            (N,17)
            (D,6)
            (SD,1)
        };
        \end{axis}
        \end{tikzpicture}
        &
        % Chart c
        \begin{tikzpicture}
        \begin{axis}[
            ybar,
            ylabel={Participants},
            symbolic x coords={SA, A, N, D, SD},
            xtick=data,
            xticklabel style={rotate=0},
            nodes near coords,
            ymin=0,
            ymax=20,
            bar width=10pt,
            enlarge x limits=0.2,
            width=0.28\textwidth,
            height=4cm,
        ]
        \addplot coordinates {
            (SA,10)
            (A,17)
            (N,10)
            (D,3)
            (SD,0)
        };
        \end{axis}
        \end{tikzpicture}
        \\
        \textbf{a)} & \textbf{b)} & \textbf{c)}
    \end{tabular}
    \end{adjustbox}
    \caption{Participant agreement levels: (a) The best way to use AI tools for improving productivity is to break a problem into small chunks and ask AI to implement it; (b) As code size and code complexity increase, productivity gains decrease because AI tools take a longer time to process and return the larger code segments; (c) As code size and code complexity increase,  productivity gains decrease because AI tools are more likely to break existing code. Response scale: SA = Strongly Agree, A = Agree, N = Neutral, D = Disagree, SD = Strongly Disagree.}
    \label{figure:productivityFactors}
\end{figure}

The overwhelming majority of the survey participants agree that the best way to use AI tools to improve productivity is to break a problem into small chunks and ask AI to implement it (Figure \ref{figure:productivityFactors}-a). As indicated in Figure \ref{figure:productivityFactors}-b, 40\% survey participants observed increase in processing time of AI tools as their code size increases, affecting productivity in a negative way. Most of the survey participants (more than 67\%) observed that AI tools are more likely to break the existing code, hindering the productivity, as the size of the code base increases (Figure \ref{figure:productivityFactors}-c).

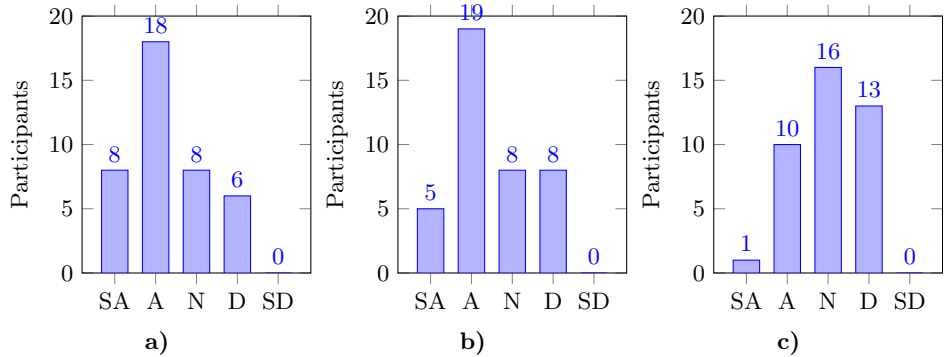
\begin{figure}[htbp]
    \centering
    \begin{adjustbox}{max width=\textwidth}
    \begin{tabular}{ccc}
        % Chart a
        \begin{tikzpicture}
        \begin{axis}[
            ybar,
            ylabel={Participants},
            symbolic x coords={SA, A, N, D, SD},
            xtick=data,
            xticklabel style={rotate=0},
            nodes near coords,
            ymin=0,
            ymax=20,
            bar width=10pt,
            enlarge x limits=0.2,
            width=0.28\textwidth,
            height=5cm,
        ]
        \addplot coordinates {
            (SA,8)
            (A,18)
            (N,8)
            (D,6)
            (SD,0)
        };
        \end{axis}
        \end{tikzpicture}
        &
        % Chart b
        \begin{tikzpicture}
        \begin{axis}[
            ybar,
            ylabel={Participants},
            symbolic x coords={SA, A, N, D, SD},
            xtick=data,
            xticklabel style={rotate=0},
            nodes near coords,
            ymin=0,
            ymax=20,
            bar width=10pt,
            enlarge x limits=0.2,
            width=0.28\textwidth,
            height=5cm,
        ]
        \addplot coordinates {
            (SA,5)
            (A,19)
            (N,8)
            (D,8)
            (SD,0)
        };
        \end{axis}
        \end{tikzpicture}
        &
        % Chart c
        \begin{tikzpicture}
        \begin{axis}[
            ybar,
            ylabel={Participants},
            symbolic x coords={SA, A, N, D, SD},
            xtick=data,
            xticklabel style={rotate=0},
            nodes near coords,
            ymin=0,
            ymax=20,
            bar width=10pt,
            enlarge x limits=0.2,
            width=0.28\textwidth,
            height=5cm,
        ]
        \addplot coordinates {
            (SA,1)
            (A,10)
            (N,16)
            (D,13)
            (SD,0)
        };
        \end{axis}
        \end{tikzpicture}
        \\
        \textbf{a)} & \textbf{b)} & \textbf{c)}
    \end{tabular}
    \end{adjustbox}
    \caption{
        Participant agreement levels on AI limitations for code understanding and integration:
        (a) Using AI to make code changes can be more time-consuming than doing it manually;
        (b) AI can take significantly longer to fix bugs than humans; (c) AI-generated code requires significant integration effort.
        Response scale: SA = Strongly Agree, A = Agree, N = Neutral, D = Disagree, SD = Strongly Disagree.
    }
    \label{figure:codeChangesBugFixingIntegrationEffort}
\end{figure}

Although 65\% of the survey participants agree that AI can take more time than humans to change the code (Figure \ref{figure:codeChangesBugFixingIntegrationEffort}-a), there is a strong indication from 15\% of the participants who disagree with this claim. This is an interesting result, as it shows that AI in certain scenarios can be faster than humans when it is necessary to change a code base. When it comes to fixing bugs, the results of the survey (Figure \ref{figure:codeChangesBugFixingIntegrationEffort}-b) show that 20\% of the participants experienced that AI is faster than humans. However, 60\% of the participants still experience that humans are (much) faster when it comes to fixing bugs. The survey results (Figure \ref{figure:codeChangesBugFixingIntegrationEffort}-c) show that there is a notable number of people who expressed that they required a significant effort when integrating AI-generated code into existing code. However, there are also those that consider this not to be a challenge at all (over 35\% of the surveyed participants). 

\begin{figure}[htbp]
    \centering
    \begin{adjustbox}{max width=\textwidth}
    \begin{tabular}{ccc}
        % Chart a
        \begin{tikzpicture}
        \begin{axis}[
            ybar,
            ylabel={Participants},
            symbolic x coords={SA, A, N, D, SD, NA, OH},
            xtick=data,
            xticklabel style={rotate=45, font=\scriptsize},
            nodes near coords,
            ymin=0,
            ymax=25,
            bar width=10pt,
            enlarge x limits=0.15,
            width=0.32\textwidth,
            height=5cm,
        ]
        \addplot coordinates {
            (SA,4)
            (A,13)
            (N,9)
            (D,5)
            (SD,1)
            (NA,8)
        };
        \end{axis}
        \end{tikzpicture}
        &
        % Chart b
        \begin{tikzpicture}
        \begin{axis}[
            ybar,
            ylabel={Participants},
            symbolic x coords={SA, A, N, D, SD},
            xtick=data,
            xticklabel style={rotate=45, font=\scriptsize},
            nodes near coords,
            ymin=0,
            ymax=25,
            bar width=10pt,
            enlarge x limits=0.15,
            width=0.32\textwidth,
            height=5cm,
        ]
        \addplot coordinates {
            (SA,3)
            (A,22)
            (N,11)
            (D,3)
            (SD,1)
        };
        \end{axis}
        \end{tikzpicture}
        &
        % Chart c
        \begin{tikzpicture}
        \begin{axis}[
            ybar,
            ylabel={Participants},
            symbolic x coords={SA, A, N, D, SD},
            xtick=data,
            xticklabel style={rotate=45, font=\scriptsize},
            nodes near coords,
            ymin=0,
            ymax=25,
            bar width=10pt,
            enlarge x limits=0.15,
            width=0.32\textwidth,
            height=5cm,
        ]
        \addplot coordinates {
            (SA,3)
            (A,18)
            (N,12)
            (D,7)
            (SD,0)
        };
        \end{axis}
        \end{tikzpicture}
        \\
        \textbf{a)} & \textbf{b)} & \textbf{c)}
    \end{tabular}
    \end{adjustbox}
    \caption{
        Perceived quality of AI-generated code: 
        (a) When generating full applications or large portions of code involving multiple functions/classes at once, partition and organisation of AI-generated code can be of a significantly lesser quality to that of human-written code and therefore not easy to maintain;     (b) Code snippets (smaller code chunks) generated by AI are of comparable maintenance quality (high cohesion, low coupling) to that of human-written code; (c) Code generated by AI is of comparable performance quality to that of human-written code.
        Response scale: SA = Strongly Agree, A = Agree, N = Neutral, D = Disagree, SD = Strongly Disagree, NA = Not Applicable.
    }
    \label{figure:aiCodeQualityThreeCharts}
\end{figure}
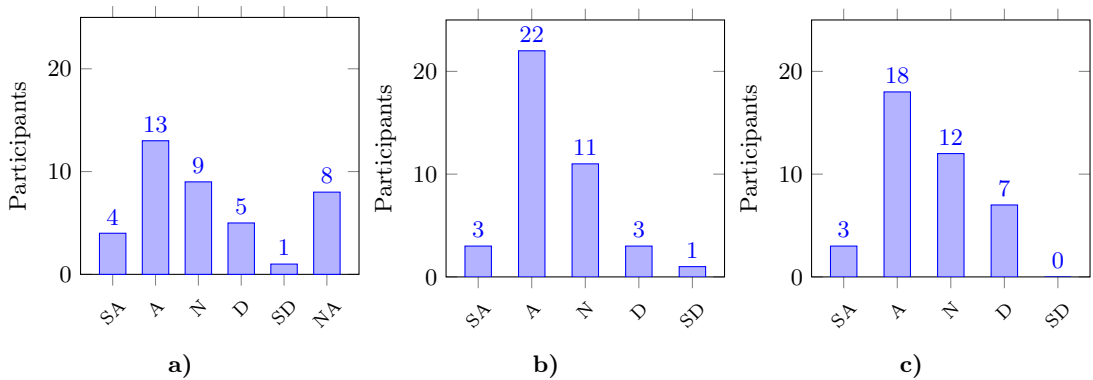

Some survey participants claim challenges with the code organization of the generated code (Figure \ref{figure:aiCodeQualityThreeCharts}-a). Although, it seems that there were few (15\%) who had positive experiences with AI partitioning the code (from the answers we had in the survey, we could not find any correlation between these results and any other answers - except that all those who answered in this way have over 3 years of experience in software engineering). The survey participants agree that generating small snippets of code with AI tool and their adoption does not lead to architecture erosion in terms of cohesion, coupling, and logical code partitioning (Figure \ref{figure:aiCodeQualityThreeCharts}-b). It seems that over 50\% of the survey participants agree that AI generated code is performant (no significant unnecessary overheads in the code that humans would avoid) in a similar way that it would have been if it was written by humans (Figure \ref{figure:aiCodeQualityThreeCharts}-c). It is important to emphasize here that 17.5\% of the survey participants have different experience and disagree with this claim, indicating that AI tools still have some work to do when it comes to code performance. However, from the answers we gathered, we could not find any correlation with other parameters that would indicate why this is the case (similar tools, years of experience, company size, and roles as for those that have experienced the opposite).

% --- Second Figure ---

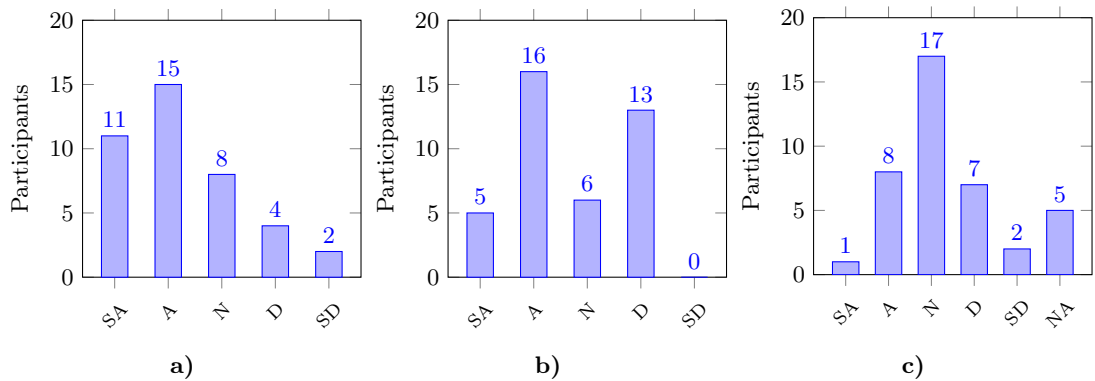
\begin{figure}[htbp]
    \centering
    \begin{adjustbox}{max width=\textwidth}
    \begin{tabular}{ccc}
        % Chart a - Functional adequacy
        \begin{tikzpicture}
        \begin{axis}[
            ybar,
            ylabel={Participants},
            symbolic x coords={SA, A, N, D, SD},
            xtick=data,
            xticklabel style={rotate=45, font=\scriptsize},
            nodes near coords,
            ymin=0,
            ymax=20,
            bar width=10pt,
            enlarge x limits=0.15,
            width=0.32\textwidth,
            height=5cm,
        ]
        \addplot coordinates {
            (SA,11)
            (A,15)
            (N,8)
            (D,4)
            (SD,2)
        };
        \end{axis}
        \end{tikzpicture}
        &
        % Chart b - Syntactic correctness
        \begin{tikzpicture}
        \begin{axis}[
            ybar,
            ylabel={Participants},
            symbolic x coords={SA, A, N, D, SD},
            xtick=data,
            xticklabel style={rotate=45, font=\scriptsize},
            nodes near coords,
            ymin=0,
            ymax=20,
            bar width=10pt,
            enlarge x limits=0.15,
            width=0.32\textwidth,
            height=5cm,
        ]
        \addplot coordinates {
            (SA,5)
            (A,16)
            (N,6)
            (D,13)
            (SD,0)
        };
        \end{axis}
        \end{tikzpicture}
        &
        % Chart c - Complex testing techniques
        \begin{tikzpicture}
        \begin{axis}[
            ybar,
            ylabel={Participants},
            symbolic x coords={SA, A, N, D, SD, NA},
            xtick=data,
            xticklabel style={rotate=45, font=\scriptsize},
            nodes near coords,
            ymin=0,
            ymax=20,
            bar width=10pt,
            enlarge x limits=0.15,
            width=0.32\textwidth,
            height=5cm,
        ]
        \addplot coordinates {
            (SA,1)
            (A,8)
            (N,17)
            (D,7)
            (SD,2)
            (NA,5)
        };
        \end{axis}
        \end{tikzpicture}
        \\
        \textbf{a)} & \textbf{b)} & \textbf{c)}
    \end{tabular}
    \end{adjustbox}
    \caption{
        Participant responses on characteristics of AI-generated code:
        (a) AI-generated code rarely fulfils desired functionality in the first shot. As the problem size increases, so does the need for fixing the generated code, 
        (b) AI-generated code is almost always syntactically correct, 
        (c) Tests generated by AI tend to use more complex testing techniques (e.g., mock objects, pixel colour testing).
        Response scale: SA = Strongly Agree, A = Agree, N = Neutral, D = Disagree, SD = Strongly Disagree, NA = Not Applicable.
    }
    \label{figure:ai-code-functional-syntax-testing}
\end{figure}

It seems to be a common experience in industry that first shot answers provided by AI tools rarely fulfill desired functionality, and therefore prompts have to be tuned (Figure \ref{figure:ai-code-functional-syntax-testing}-a). Over 50\% of the survey participants say that AI-generated code is almost always syntactically correct. But, there is a significant number of participants who disagree (Figure \ref{figure:ai-code-functional-syntax-testing}-b). However, from the answers we gathered, we could not find any correlation with other parameters that would indicate why this is the case (similar tools, years of experience, company size, and roles as for those that have experienced the opposite). When it comes to the tests, some survey participants agree that AI generated tests use more complex testing techniques (around 20\%), while the same number of the survey participants disagree with this claim. The majority was neutral with respect to this question (Figure \ref{figure:ai-code-functional-syntax-testing}-e).

\begin{figure}[ht]
\centering

% --- First Chart: Productivity Impact ---
\begin{minipage}[t]{0.48\textwidth}
\centering
\begin{tikzpicture}
\begin{axis}[
    ybar,
    bar width=0.4cm,
    width=0.95\textwidth,
    height=4.5cm,
    ylabel={Participants},
    symbolic x coords={
        {Much higher},
        {Higher},
        {Around 35\%},
        {Lower},
        {Zero}
    },
    xtick=data,
    x tick label style={align=center, rotate=45},
    nodes near coords,
    nodes near coords align={vertical},
    ymin=0,
    enlarge y limits=false,
    axis on top=true
]
\addplot coordinates {
    ({Much higher},1)
    ({Higher},17)
    ({Around 35\%},14)
    ({Lower},7)
    ({Zero},1)
};
\end{axis}
\end{tikzpicture}
\caption{Perceived productivity improvements from AI tools considering the reference 35\% improvement}
\label{figure:productivityImprovements}
\end{minipage}
\hfill
% --- Second Chart: Best Moment to Apply AI Tools ---
\begin{minipage}[t]{0.48\textwidth}
\centering
\begin{tikzpicture}
\begin{axis}[
    ybar,
    bar width=0.4cm,
    width=0.95\textwidth,
    height=4.5cm,
    ylabel={Participants},
    symbolic x coords={
        {Strongly\\agree},
        Agree,
        Neutral,
        Disagree,
        {Strongly\\disagree}
    },
    xtick=data,
    x tick label style={align=center, rotate=45},
    nodes near coords,
    nodes near coords align={vertical},
    ymin=0,
    enlarge y limits=false,
    axis on top=true
]
\addplot coordinates {
    ({Strongly\\agree},6)
    (Agree,18)
    (Neutral,7)
    (Disagree,6)
    ({Strongly\\disagree},1)
};
\end{axis}
\end{tikzpicture}
\caption{The best moment to apply AI tools is in the early code implementation stages}
\label{figure:momentToApplyAITools}
\end{minipage}

\end{figure}

When it comes to \textbf{individual comments} we received about \textbf{positive effects of AI tools on productivity}, the survey participants listed several interesting statements (we tried to merge similar ones). For example: i) use LLMs to quickly sum up and recommend the latest cutting-edge research that’s relevant to the daily work, which really speeds up getting new tech into software. They claim that this has been more useful than help with coding; ii) Quickly access information from design documents and other company-specific resources (much faster than standard searching); iii) Brainstorm ideas, ask questions like "how do I do this?" rather than asking the tool to output code; iv) Ask AI tools for example implementation with some libraries is faster than reading documentation; v) Generate boilerplate code.

\subsection{Summary}\label{sec:lessonsLearned:architecture}

%In the test project, we have established that AI tools can have an influence on software architecture. This is reflected and the adoption of generated code and in cases when we ask AI to suggest architectural solutions. Using AI suggested solutions can have its benefits (as we saw in the test project, AI generated architecture predicted some new features that then became quite easy to implement and outperformed a human). However, there are also potential drawbacks (integration between parts designed by AI can be challenging). 
Here are the conclusions based on the survey analysis:

i) About 45\% of the survey participants experienced an increase in productivity of OVER 35\% when using AI tools. This fact cannot be ignored, and it is clear that anyone working today in the software industry needs to use AI tools if they want to remain competitive. Otherwise, with the best of their efforts, they will struggle to keep up with the productivity that those who use AI tools can achieve.

ii) Adoption of AI-generated code snippets does NOT lead to a significant erosion of software architecture (in terms of cohesion, coupling, partitioning of logical parts). The code is performant and easy to maintain. AI-generated code is almost always syntactically correct and the quality of the logical syntax style is adequate. 

iii) AI tools can suggest adequate architectural designs. However, this conclusion should be taken carefully, as it depends on many factors, including how we formulate the problem and the experience of those that consume the AI generated solutions, as solutions often require tweaks. Junior software architects would certainly struggle to assess whether AI-generated solutions are adequate. %, but AI can give them a good pointer in which direction to go. % and do research on their own.

iv) However, AI tends to suggest inadequate solutions when faced directly with large and complex problems (without prior decomposition). In addition, AI tends to suggest inadequate solutions when changing complex and large code bases. In such cases, partition and organisation of AI-generated code is often of a significantly lesser quality than that of human-written code, and therefore not easy to maintain and integrate. Furthermore, using AI to change complex and large code bases leads to the breaking of existing code.

Although engineers tend to profit from AI in terms of higher productivity, it is also clear that they still need to perform a decomposition of problems to the level that AI can handle. Furthermore, it is also clear that the job of software engineers is to synthesise the generated solutions into well-architected systems. Therefore, it is fair to say that architecture skills are highly needed in the age of AI.

\section{Conclusion and Future Work}\label{sec:conclusion}

%There are many challenges with adopting AI tools in software engineering. 
Generating code and tests is only one of the use cases where the adoption of AI tools increases productivity. Others include using AI for more conceptual work (software architecture) and using AI to help with side tasks (e.g., summarising information). Despite all the challenges associated with the use of AI in software engineering, it is clear that AI is not a hype and that AI tools in software engineering are here to stay. The increase in productivity of over 35\% is a clear and strong sign that individuals using AI tools significantly outperform those who do not use AI tools. The questions that now open are related to the use cases where we could apply AI, how to reduce friction when adopting AI tools in existing workloads and code bases, and what is it that we can now achieve with the increase in productivity. It is also clear that not everyone equally enjoys increases in productivity, and there is a strong concern that junior engineers might bring more overhead than benefit when adopting solutions that they fully do not understand.

As code size and code complexity increase (in general, as a problem complexity increases), productivity gains decrease because AI tools are a) more likely to break existing code; b) demonstrate increased processing time; c) software shows the sign of architectural erosion. Therefore, it is more challenging to change code using AI than to create new code. We have to observe that today this is where most of the engineering time is invested, and it would be great if AI could support these activities in a smoother way. That will also be one of the topics of our future work.

\bibliographystyle{splncs04}
\bibliography{references}

\end{document}